\documentclass{appolb}
\usepackage{graphicx}

\def\half{{\textstyle{1\over2}}}

\def\CG#1#2#3#4#5#6{C^{#5#6}_{#1#2#3#4}}
\newcommand{\bi}[1]{\hbox{\boldmath{$#1$}}}%

\begin{document}

\title{STUDIES OF THE ROPER RESONANCE BY THE LJUBLJANA GROUP}

\author{Simon \v{S}irca
\address{Faculty of Mathematics and Physics, University of Ljubljana, Slovenia}
\address{Jo\v{z}ef Stefan Institute, Ljubljana, Slovenia}}

\maketitle

\begin{abstract}
Ever since its discovery in 1964 the nature of the $N^\ast(1440)$ 
nucleon resonance has been a perpetual and one of the outstanding 
puzzles in hadronic physics.
The Ljubljana group joined the global effort in the late 1990s, first 
from the theoretical viewpoint and later experimentally.  This paper is 
a short overview of our attempts to understand this elusive resonance.
\end{abstract}

\section{Introduction}

The first excited state of the nucleon with the same quantum
numbers, the $N^\ast(1440)$ or the ``Roper'' resonance \cite{Roper:1964zza}, 
has evaded detailed understanding by the hadronic community for decades.
One of the reasons is that it is anything but conspicuous, being hidden
below much more prominent neighboring resonances, in particular
the $\Delta(1232)$ ($P_{33}$) and the $N^\ast(1520)$ ($D_{13}$) which 
were already known at the time of the discovery, possessing an unusually 
large width and exhibiting a very atypical behavior of $\mathrm{Im}\,T_{\pi N}$ 
in the $P_{11}$ partial wave.  Indeed Professor Roper himself appeared 
to be reluctant to accept the rather peculiar feature in the Argand 
diagram and, as quoted in \cite{Alvarez-Ruso:2010ayw}, lamented: 
``I spent [a] much time trying to eliminate the $P_{11}$ resonance.'' 
The progress brought about by partial-wave analyses was not as fast
as one would hope for and did not converge well, and I can still 
recall Professor Arndt's response at a 2004 conference at ACU 
to a question on how one could best improve the determination 
of the $P_{11}$ mass(es?), width, pole position and decay properties.  
It was indicative of the issues that the partial-wave community was facing: 
``I've expressed my position 
on this subject many times. It just isn't possible to fit $P_{11}$ 
with a `simple' Breit-Wigner form; the amplitude is determined by nearby 
singularities consisting of two poles and a very prominent cut 
($\pi$--$\Delta$).  It's like doing a polynomial fit to a sine wave. 
[...]~I believe that the `problem' with $P_{11}$ is that people keep 
trying to stuff a square pole in a round hole.''

As members of two hadron-physics groups from the Jo\v{z}ef Stefan 
Institute and the Faculty of Mathematics and Physics in Ljubljana, 
Slovenia, consisting of Professor Mitja Rosina, late Professor 
Bojan Golli and the Author,
we were motivated to investigate the Roper resonance partly due to our 
existing know-how on statical and dynamical properties of the nucleons
studied in terms of quark models.  These model calculations,
initially performed in collaboration with our Portuguese colleagues
from Coimbra (J.~da Provid\^{e}ncia, J.~Urbano, M.~Fiohais, P.~Alberto), 
were initiated even prior to that, and were thriving on an earlier
substrate of the Peierls--Yoccoz projection technique, coherent
states and the ``hedgehog'' ansatz for the baryon wave functions
\cite{PhysRevD.18.4208}.

\section{The Roper resonance in the chromodielectric model}

Our first attempt at explaining a specific feature of the Roper
resonance in terms of a quark model was to calculate the transverse 
and scalar helicity amplitudes for the nucleon--Roper transition, 
$A_{1/2}$ and $S_{1/2}$.  The Roper resonance was described in
a chiral version of the chromodielectric model (CDM) as consisting
of a three-quark core with one of the quarks promoted to the $2s$ 
orbit, surrounded by a cloud of pions and $\sigma$-mesons and
furnished with a chromodielectric field $\chi$ which ensures that 
the quarks remain dynamically confined \cite{Alberto:2001fy}.
The $\chi$-field is a peculiar ingredient of the CDM, and this
feature of the model has also been relevant in other contexts,
in particular in quantifying the role of the pion cloud, 
that is, ``counting pions in the nucleon'' \cite{FIOLHAIS1996151}.

The wave functions of the physical nucleon and the Roper resonance
were obtained by performing a Peierls--Yoccoz projection on the
baryon wave function in the form of a ``hedgehog'' coherent state
and making them orthogonal to one another.  The Roper resonance 
was represented as a breathing mode of the three valence quarks 
in the bare core (the source), with all three meson fields adapting 
to the changes 
in the source; all fields were computed in a self-consistent manner.  
(Alternatively, the Roper could be modeled by retaining all three
core quarks in the ground state but letting the $\chi$-field
and/or the $\sigma$-field oscillate.)  We have computed the
$Q^2$-dependence of the helicity amplitudes $A_{1/2}^{p}$
and $S_{1/2}^{p}$ in the range $0 \le Q^2 \le 2\,(\mathrm{GeV}/c)^2$
and obtained correct signs and reasonable trends at a time when
the world data consisted merely of three points with unspecified 
(possibly huge) uncertainties which, moreover, were at odds with
the rather accurately determined helicity couplings at $Q^2=0$.
We also computed $A_{1/2}^{n}(Q^2)$ for which the only
experimental reference was at the photon point, missing it 
by a factor of approximately two.

\section{Coupled-channel approach with chiral quark models}

Because in the CDM the $\chi$-field takes care of the confinement,
the accompanying pion field is accordingly weaker, resulting in a smaller 
average 
number of pions and, we presumed, in a relatively minor contribution
of the pion cloud in the photo-excitation amplitudes.  The abundance 
of theoretical and experimental studies of the $N$--$\Delta$ transition 
at the turn of the millennium---see, for instance, \cite{AZNAURYAN20121} 
and references therein for a review---provided ample evidence that
the pion cloud indeed plays an important role, especially at small $Q^2$.
This motivated us to open a much wider front and to construct 
a coupled-channel framework that would allow for the analysis
of the electro-magnetic excitation parts of the photo- and electro-production
 processes as well as their strong decay parts.  The framework that
 we devised \cite{Golli:2007sa} is in principle
able to accommodate any underlying quark model, but we have chosen to
perform all our calculations by using the Cloudy Bag Model (CBM),
for the simple reason that the quark-meson interaction matrix elements
were relatively easy to compute, and in addition due to its presumed 
advantage of featuring a much stronger pion cloud than, say, the CDM.

In our coupled-channel $K$-matrix formalism we only consider models
in which the the meson fields couple linearly to the quark core,
and there is no meson self-interaction.  The part of the Hamiltonian
corresponding to mesons has the form
\begin{equation}
   H_\mathrm{m} = 
     \int\mathrm{d}k \sum_{lmt}
   \left\{\omega_k\,a^\dagger_{lmt}(k)a_{lmt}(k) +
         \left[{V}_{lmt}(k) a_{lmt}(k)
             + {V}_{lmt}(k)^\dagger\,a^\dagger_{lmt}(k)\right] \right\} \>,
\label{Hm}
\end{equation}
where $a^\dagger_{lmt}(k)$ is the creation operator for an $l$-wave
meson with the third component of spin $m$ and---in the case of isovector 
mesons---the third component of isospin $t$.
In the case of the $p$-wave pions, the source can be cast 
in the form
\begin{equation}
      V_{1mt}(k) = -v(k)\sum_{i=1}^3 \sigma_m^i\tau_t^i\>,
\label{Vmt}
\end{equation}
with $v(k)$ depending on a particular quark model and containing 
the information about the underlying quark structure.  Note that
${V}_{lmt}(k)$ may also induce radial excitations of the quark core,
in particular $1s\rightarrow 2s$, $1s\rightarrow 1p_{1/2}$ and
 $1s\rightarrow 1p_{3/2}$ transitions.  In the model of choice, CBM,
the relevant operators are
\begin{eqnarray*}
  {V_{1mt}^{s\to s}(k)} &=& 
    {1\over2{f}}\, {k^2\over\sqrt{12\pi^2\omega_k}}\,
    {\omega_{s}\over\omega_{s}-1}\,
            {j_1(k{R_\mathrm{bag}})\over  k{R_\mathrm{bag}}}
           \sum_{i=1}^3 {\sigma_m^i\tau_t^i}\>, \\[2pt]
  {V_{1mt}^{s\to 2s}(k)} &=& 
    {1\over2{f}}\, {k^2\over\sqrt{12\pi^2\omega_k}}\,
    \sqrt{{\omega_{2s} \omega_\mathrm{s}\over
       (\omega_{2s}-1)(\omega_\mathrm{s}-1)}}\,
            {j_1(k{R_\mathrm{bag}})\over  k{R_\mathrm{bag}}}
           \sum_{i=1}^3 {\sigma_m^i\tau_t^i}\>, \\[2pt]
  {V_{00t}^{s\to p_{1/2}}(k)} &=& 
    {1\over2{f}}\, {k^2\over\sqrt{4\pi^2\omega_k}}\,
    \sqrt{{\omega_{{p}_{1/2}}\omega_{s}\over
        (\omega_{{p}_{1/2}}+1)(\omega_{s}-1)}}\,
            {j_0(k{R_\mathrm{bag}})\over  k{R_\mathrm{bag}}}
           \sum_{i=1}^3 {\tau_t^i}\>, \\[2pt]
  {V_{2mt}^{s\to p_{3/2}}(k)} &=& 
    {1\over2{f}}\, {k^2\over\sqrt{2\pi^2\omega_k}}\,
    \sqrt{{\omega_{{p}_{3/2}}\omega_{s}\over
        (\omega_{{p}_{3/2}}-2)(\omega_{s}-1)}}\,
            {j_2(k{R_\mathrm{bag}})\over  k{R_\mathrm{bag}}}
           \sum_{i=1}^3 {\Sigma_{2m}^i\tau_t^i}\>,
\end{eqnarray*}
involving $p$-wave, $p$-wave, $s$-wave and $d$-wave pions, respectively.

The next step is to construct the $K$ matrix.  The idea is to include 
many-body states of quarks (and mesons) in the scattering formalism 
in a Chew-Low type approach \cite{PhysRev.101.1570}.  This is most
easily accomplished in the $JT$ basis in which the $K$ and $T$ matrices 
are diagonal.  The $K$-matrix has the form
\begin{equation}
   {K}_{\pi N\pi N}^{{JT}}(k,k_0) =  -\pi\sqrt{\omega_k E_N\over k\, W}
      {\langle\Psi^N_{JT}(W)}||V(k)||\Psi_N\rangle \>,
\label{Kmat1}
\end{equation}
and we introduce the principal-value states
\begin{equation}
 {|\Psi^N_{JT}(W)\rangle} = \sqrt{\omega_0E_N\over k_0 W}
  \left\{
   \left[a^\dagger(k_0)|\Psi_N\rangle\right]^{{JT}}
  - {\mathcal{P}\over H-W}\,
  \left[V(k_0)|\Psi_N\rangle\right]^{{JT}}
    \right\} \>,
\label{pvstate}
\end{equation}
where $[~~~]^{JT}$ denotes coupling to good $J$ and $T$, 
$W$ is the invariant energy of the system, $\omega_0$ and $k_0$ 
are the energy and momentum of the pion, and $E_N$ is the nucleon energy.

\begin{figure}[!htbp]
\centerline{%
\includegraphics[width=10cm]{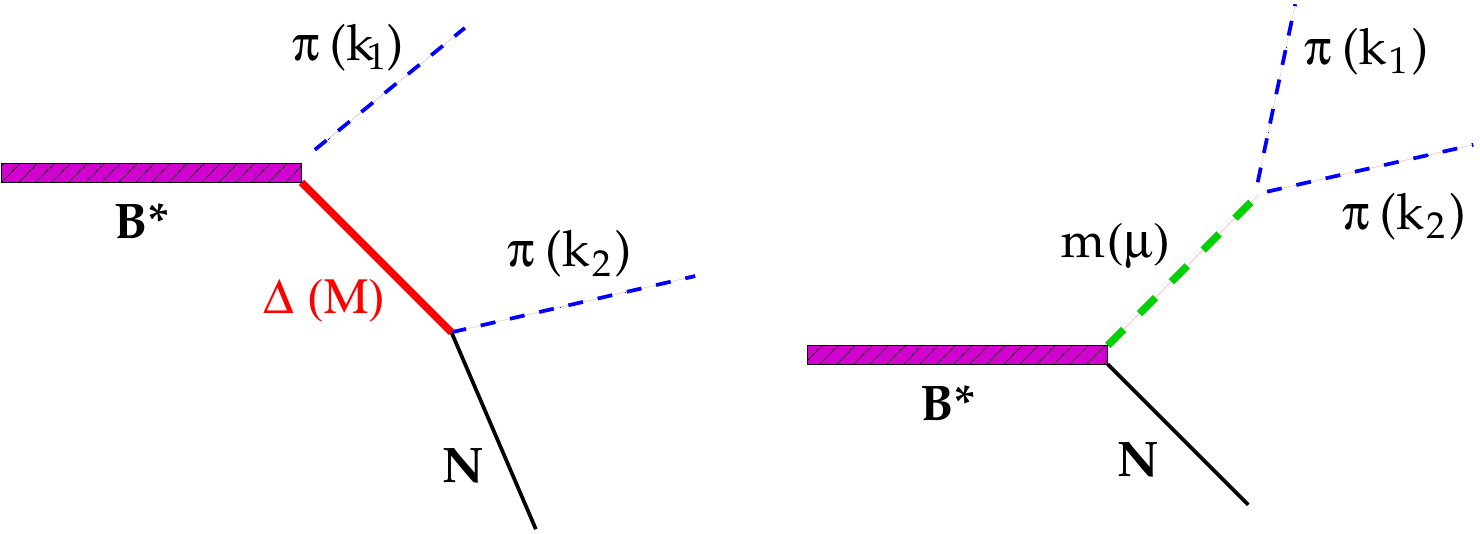}}
\caption{Two-pion decay of an excited baryon involving  
an intermediate unstable baryon with invariant mass $M$
like the $\Delta$ (left) or an unstable meson with invariant
mass $\mu$, for instance, the $\sigma$-meson (right).}
\label{fig:2piDecay}
\end{figure}

This set of formulas has then been extended to the multi-channel case,
assuming that two-pion decays proceed through some intermediate
unstable particle, either a baryon or a meson, as shown in 
Fig.~\ref{fig:2piDecay}.  The task at hand, then, was to devise
principal-value states with suitable orthogonality and normalization
properties, and use them to construct a multi-channel $K$-matrix 
with the general form
\begin{center}
\begin{tabular}{|l|l|l|}
\hline
$K_{NN}$            & $K_{NB}(M')$        & $K_{N\sigma}(\mu')$ \\
\hline
$K_{B N}(M)$   & $K_{BB'}(M,M')$ & $K_{B\sigma}(M,\mu')$ \\
\hline
$K_{\sigma N}(\mu)$ & $K_{\sigma B}(\mu,M)$ 
 & $K_{\sigma\sigma}(\mu,\mu')$ \\
\hline
\end{tabular}
\end{center}
Once the $K$-matrix is obtained, the $T$ and $S$ matrix 
are given by the relations
\begin{equation}
    T = {K\over 1-\mathrm{i} K}\>, \qquad
    S = 1 + 2\mathrm{i}T\>.
\label{TSmatrix}
\end{equation}
In integral form, the first equation in (\ref{TSmatrix}),
$T = K + \mathrm{i}\,TK$, amounts to the set of coupled integral 
(Heitler) equations:
\begin{eqnarray}
T_{NN} &=& K_{NN} + \mathrm{i} T_{NN}K_{NN} \nonumber \\
       &+& \mathrm{i}\sum_B\int_{M_N+m_\pi}^{W-m_\pi}\mathrm{d} M\,T_{NB}(M) 
                            K_{B N}(M) \nonumber \\
       &+& \mathrm{i}\int_{2m_\pi}^{W-M_N}\mathrm{d}\mu\,T_{N\sigma}(\mu) 
                            K_{\sigma N}(\mu)\>, \nonumber \\
T_{N B}(M) &=& K_{N B}(M)  +\mathrm{i} T_{NN}K_{NB}(M) \nonumber \\
      &+& \mathrm{i}\sum_{B'}\int_{M_N+m_\pi}^{W-m_\pi}\mathrm{d} M'\,T_{NB'}(M') K_{B'B}(M',M) \label{eq4T} \\
      &+& \mathrm{i}\int_{2m_\pi}^{W-M_N}\mathrm{d}\mu\,T_{N\sigma}(\mu) 
         K_{\sigma B}(\mu,M)\>, \nonumber\\
T_{N \sigma}(\mu) &=& K_{N \sigma}(\mu)  +\mathrm{i} T_{NN}K_{N\sigma}(\mu) \nonumber \\
   &+& \mathrm{i}\sum_{B}\int_{M_N+m_\pi}^{W-m_\pi}\mathrm{d} M\,T_{NB}(M) 
         K_{B\sigma}(M,\mu) \nonumber \\ 
   &+& \mathrm{i}\int_{2m_\pi}^{W-M_N}\mathrm{d}\mu'\,T_{N\sigma}(\mu') 
         K_{\sigma\sigma}(\mu',\mu)\>. \nonumber
\end{eqnarray}

The desired principal-value states $|\Psi_{JT}^B(W,M)\rangle$ 
corresponding to the $\pi B$ channel (where $B$ stands for any 
intermediate baryon, for instance, the $\Delta$) were introduced 
in a form analogous to (\ref{pvstate}):
\begin{equation}
 |\Psi^B_{JT}(W,M)\rangle = \mathcal{N}_1
  \biggl\{
   \left[a^\dagger(k_1)|\widetilde{\Psi}_B(M)\rangle\right]^{JT}
  \!\! - {\mathcal{P}\over H\!-\!W}\,
  \left[V(k_1)|\widetilde{\Psi}_B(M)\rangle\right]^{JT}
    \biggr\} \>,
\label{pvstate2}
\end{equation}
where $~\widetilde{}~$ signifies that orthonormality has been
accounted for.  The $K$-matrix elements computed 
with such states are
\begin{eqnarray*}
   K_{\pi B\pi N}^{JT}(k,k_0,M) 
  &=&  -\pi\sqrt{\omega_k E\over k\, W}
     \langle\Psi^N_{JT}(W)||V(k)||
                          \widetilde{\Psi}_B(M)\rangle\>, \\
   K_{\pi N\pi B}^{JT}(k,k_1,M) &=&  -\pi\sqrt{\omega_k E\over k\,W}
     \langle\Psi^B_{JT}(W,M)||V(k)||\Psi_N\rangle \>, \\
K_{\pi B'\pi B}^{JT}(k,k_1,M',M) &=&
     -\pi\sqrt{\omega_k E\over k\, W}
   \langle\Psi^B_{JT}(W,M)||V(k)||\tilde{\Psi}_{B'}(M')\rangle \>,
\label{KCL-BB}
\end{eqnarray*}
describing, from top to bottom, a process in which the initial
$\pi N$ system with invariant mass $W$ decays into a pion and 
a $\pi N$ system with the quantum numbers of the intermediate 
baryon $B$; the $\pi B\to \pi N$ transition; and the 
$\pi B\to \pi B'$ transition, for instance,
$\pi(k_1)+\Delta(M)\rightarrow \pi(k)+\Delta(M')$.
Similar expressions are derived for the channels
involving the $\sigma$-mesons.  (For purposes of brevity
and in tune with the scope of the paper the lengthy expressions 
with similar structure for different inelastic channels shall
also be omitted in the following; consult
\cite{Golli:2007sa} for details.)

One of the obstacles encountered was that in formal expressions 
like (\ref{pvstate}) or (\ref{pvstate2}) the interaction
$V(k)$ generates bare three-quark states with quantum numbers 
different from the ground state, as well as superpositions of bare 
three-quark states dressed with mesons.  This problem has been
resolved by exploiting general expressions for the matrix elements
of the pion field between the eigenstates of the Hamiltonian
whose pion part has the form (\ref{Hm}); see Appendix A 
of \cite{Golli:2007sa}.  As an illustration, the elastic channel
has been represented by the ansatz
\begin{eqnarray}
|\Psi_{JT}^N(W)\rangle 
 &=& \mathcal{N}_0\,
\biggl\{
     \sum_\mathcal{R} c_\mathcal{R}^N(W)|\Phi_\mathcal{R}\rangle
   + [a^\dagger(k_0)|\Psi_N(k_0)\rangle]^{JT} 
\biggr. \nonumber \\ && \biggl. 
\kern -16pt
   + \int\mathrm{d} k\,{\chi_{JT}^{NN}(k,k_0)\over\omega_k+E_N(k)-W}\,
      [a^\dagger(k)|\Psi_N(k)\rangle]^{JT}
\biggr. \nonumber \\ && \biggl. 
\kern -16pt
   + \sum_{B}\int\mathrm{d} M\kern-3pt\int\mathrm{d} k\,
     {\chi_{JT}^{BN}(k,k_0,M)\over\omega_k+E_{B}(k)-W}
      [a^\dagger(k)|\widetilde{\Psi}_{B}(M)\rangle]^{JT}
\biggr. \nonumber \\ && \biggl. 
\kern -16pt
   + \int\mathrm{d} \mu \int\mathrm{d} k\,
       {\chi_{JT}^{\sigma N}(k,k_0,\mu)\over\omega_{\mu k}+E_{JT}(k)-W}\,
       b^\dagger(k)|\widetilde{\Psi}_{JT}\rangle
   \biggr\}\>
\label{JTNW}
\end{eqnarray}
with the following structure: the first term is the sum over bare 
quark states, denoted by $\Phi_\mathcal{R}$; the second term corresponds 
to the unmodified (free) pion and defines the channel as well as 
determines the normalization; the three integral terms incorporating
the meson field amplitudes $\chi$ correspond to the meson-cloud contributions, 
specifically, to the one-pion state superposed on the ground state, 
to one-pion states around different excited states, and to the 
one-$\sigma$ state around either the nucleon or the $\Delta$, respectively.  
In an analogous fashion one can then construct the principal-value states 
for the inelastic channels, in particular $|\Psi_{JT}^{B}(W,{{M}})\rangle$
for the $\pi B$ channels, and $|\Psi_{JT}^{\sigma}(W,{\mu})\rangle$
for the $\sigma N$ channel.

In the channels including the pion and unstable isobars, the meson 
amplitudes $\chi$ above the one- and  two-pion thresholds are related 
to the elastic and inelastic elements of the $K$ matrix,
\begin{eqnarray}
K_{NN}(W)      &=& \pi\,\mathcal{N}_0^2\,\chi_{JT}^{NN}(k_0,k_0)\>,
\nonumber\\
K_{BN}(W,M) &=& \pi\,\mathcal{N}_0\mathcal{N}_1\,
             \chi_{JT}^{BN}(k_1,k_0,M)\>, \label{chi2KHN} \\
K_{\sigma N}(W,\mu) &=& \pi\,\mathcal{N}_0\mathcal{N}_\mu\,
             \chi_{JT}^{\sigma N}(k_\mu,k_0,\mu)\>, \nonumber
\end{eqnarray}
while above the two-pion threshold the relations read
\begin{equation}
  K_{H'H}(W,m_H',m_H)   =
\pi\,\mathcal{N}_{H'}\mathcal{N}_H\,
             \chi_{JT}^{H'H}(k_{H'},k_H,m_H',m_H) \>.
\label{chi2K}
\end{equation}
Here $H$ stands for either the $\pi B$ or $\sigma N$ channels,
and $m_H$ is the corresponding invariant mass, $M$ or $\mu$,
respectively.  

Next the equations for the pion amplitudes $\chi$, the coefficient
$c_{\cal R}$ in the ansatz (\ref{JTNW}) and the analogous coefficients
$\widehat{c}_{\cal R}^B$ appearing in $|\Psi_{JT}^{B}(W,{{M}})\rangle$  
and $\widehat{c}_{\cal R}^\sigma$ appearing in 
$|\Psi_{JT}^{\sigma}(W,{\mu})\rangle$, must be obtained.  
The appropriate equations can be derived by invoking 
the Kohn variational principle
$\langle\delta\Psi^\mathrm{P}|H-W|\Psi^\mathrm{P}\rangle = 0$,
where  $\Psi^\mathrm{P}$ is a trial state, and requiring stationarity 
with respect to the variation of the coefficients $c_\mathcal{R}^N$, 
$\widetilde{c}_\mathcal{R}^B$ and $\widetilde{c}_\mathcal{R}^\sigma$.
This procedure results in the set of equations
\begin{eqnarray}
(W-M_\mathcal{R}^0)c_\mathcal{R}^N(W) 
&=&  V_{N\mathcal{R}}(k_0) + \int\mathrm{d}k\,{V_{N\mathcal{R}}(k)\,
      \chi_{JT}^{NN}(k,k_0)
                 \over \omega_k+E_N(k)-W }
\nonumber\\ && 
+  \sum_{B'}\int\mathrm{d}k\,{ V_{B' \mathcal{R}}^{M_{B'}}(k)\,
       \widehat{\chi}_{JT}^{B' N}(k,k_0,M_{B'})
        \over \omega_k+E_{B'}(k)-W}
\nonumber\\ && 
+    \int\mathrm{d}k\,{ V_{N \mathcal{R}}^{m_\sigma}(k)\,
       \widehat{\chi}_{JT}^{\sigma N}(k,k_0,m_\sigma)
        \over \tilde{\omega}_k+ E_{JT}(k)-W}\>,
\label{eq4cRN}
\end{eqnarray}
\begin{eqnarray}
(W-M_\mathcal{R}^0)\widehat{c}_{\mathcal{R}}^{B}(W,M)
&=&
 V_{B\mathcal{R}}^{M}(k_1) 
  + \int\mathrm{d}k\,{V_{N\mathcal{R}}(k)\,
     \widehat{\chi}_{JT}^{NB}(k,k_1,M)
          \over \omega_k+E_N(k)-W}
\nonumber\\  && 
   + \sum_{B'}\int\mathrm{d}k\,
       {V_{B' \mathcal{R}}^{M_{B'}}(k) 
        \widehat{\chi}_{JT}^{B'B}(k,k_1,M_{B'},M) 
        \over 
             \omega_k+ E_{B'}(k)-W}\,
\nonumber\\  && 
   +    \int\mathrm{d}k\,
       {V_{N \mathcal{R}}^{m_\sigma}(k) 
        \widehat{\chi}_{JT}^{\sigma B}(k,k_1,m_\sigma,M) 
        \over 
             \tilde{\omega}_k+ E_{JT}(k)-W}\>,
\label{eq4cRB}
\end{eqnarray}
\begin{eqnarray}
(W-M_\mathcal{R}^0)\widehat{c}_\mathcal{R}^\sigma(W,\mu) 
&=&  V^\mu_{\sigma\mathcal{R}}(k_\mu) 
   + \int\mathrm{d}k\,{V_{N \mathcal{R}}(k)\,
         \widehat{\chi}_{JT}^{N\sigma}(k,k_\mu,\mu)
                 \over \omega_k+E_N(k)-W}
\nonumber\\ && 
+  \sum_{B'}\int\mathrm{d}k\,{ V_{B' \mathcal{R}}^{M_{B'}}(k)\,
       \widehat{\chi}_{JT}^{B' \sigma}(k,k_\mu,M_{B'},\mu)
        \over \omega_k+E_{B'}(k)-W}
\nonumber\\ && 
+    \int\mathrm{d}k\,{ V_{N \mathcal{R}}^{m_\sigma}(k)\,
       \widehat{\chi}_{JT}^{\sigma \sigma}(k,k_\mu,m_\sigma,\mu)
        \over \tilde{\omega}_k+ E_{JT}(k)-W}\>,
\label{eq4cRs}
\end{eqnarray}
where
\begin{eqnarray*}
  V_{N\mathcal{R}}(k) &=& \langle\Phi_\mathcal{R}||V(k)||\Psi_N\rangle
            = Z_N^{-1/2}\langle\Phi_\mathcal{R}||V(k)||\Phi_N\rangle\>,
\\
  V_{B\mathcal{R}}^{M}(k) &=& 
\langle\Phi_\mathcal{R}||V(k)||\widehat{\Psi}_B(M)\rangle
    = Z_B^{-1/2}\langle\Phi_\mathcal{R}||V(k)||\Phi_B\rangle\>,
\\
  V_{N\mathcal{R}}^{\mu}(k) &=& 
\langle\Phi_\mathcal{R}||V^\mu(k)||\widehat{\Psi}_N\rangle
    = Z_N^{-1/2}\langle\Phi_\mathcal{R}||V^\mu(k)||\Phi_N\rangle\>.
\end{eqnarray*}
Here $Z_B$ is the wave-function normalization, while
the reduced matrix elements $\langle\Phi_\mathcal{R}||V(k)||\Phi_N\rangle$ 
and $\langle\Phi_\mathcal{R}||V(k)||\Phi_B\rangle$ are calculated by
using the underlying quark model of choice, for instance, the CBM
in our sample calculations.

Requiring stationarity with respect to the pion amplitudes 
leads to the Lippmann-Schwinger equation for the $K$ matrix.
The equation for the $\chi_{JT}^{NN}$ amplitude, which 
is related to the elastic part of the $K$ matrix, reads
\begin{eqnarray}
\chi_{JT}^{NN}(k,k_0)
&=& {\cal K}^{NN}(k,k_0) 
    -\sum_\mathcal{R} c_\mathcal{R}^N(W) V_{N\mathcal{R}}(k)
\nonumber\\ && 
 + \int\mathrm{d}k'\,{{\cal K}^{NN}(k,k')\,\chi_{JT}^{NN}(k',k_0)\over
                    \omega_k'+E_N(k')-W}
\nonumber\\ && 
 + \sum_B \int\mathrm{d}k'\,{{\cal K}_{M_B}^{NB}(k,k')\,
                 \widehat{\chi}_{JT}^{B N}(k',k_0,M_B)
                    \over\omega_k'+E_B(k')-W}
\nonumber\\ && 
 + \int\mathrm{d}k'\,{{\cal K}_{m_\sigma}^{N\sigma}(k,k')\,
                 \widehat{\chi}_{JT}^{\sigma N}(k',k_0,m_\sigma)
                    \over\tilde{\omega}_k'+E_{JT}(k')-W}\>,
\label{eq4chiN} 
\end{eqnarray}
where we have introduced the kernels
\begin{eqnarray*}
{\cal K}_M^{NB}(k,k')
 &=&
-\sum_{mtm't'}\langle\Psi_N(k)|
a^\dagger_{m't'}(k')
\nonumber \\ 
&\times &
\Bigl[V^\dagger_{mt}(k)
            +(\omega_k+E_N(k)-W)a_{mt}(k)\Bigr]|
      \widehat{\Psi}_B(M)\rangle \\
&\times & 
      \CG{J_B}{\half-m'}{1}{m'}{J}{\half}\CG{T_B}{\half-t'}{1}{t'}{T}{\half}
      \CG{\half}{\half-m}{1}{m}{J}{\half}\CG{\half}{\half-t}{1}{t}{T}{\half}\>.
\end{eqnarray*}
(For $B=N$, $\widehat{\Psi}_B(M)$ reduces to $\Psi_N$ and
$M$ to $M_N$.)  The pion amplitudes involving the $\pi B$ 
channels, in turn, satisfy the equation
\begin{eqnarray}
\widehat{\chi}_{JT}^{B'B}(k,k_1,M',M)\,
&=&
  {\cal K}_{M'M}^{B'B}(k,k_1)
 -\sum_\mathcal{R} \widehat{c}_\mathcal{R}^{B}(W,M) V_{B' \mathcal{R}}^{M'}(k)
\nonumber\\ 
    &+& \sum_{B''}\int\mathrm{d}k'\,
  {{\cal K}_{M'M_{B''}}^{B'B''}(k,k')\,
  \widehat{\chi}_{JT}^{B''B}(k',k_1,M_{B''},M) 
         \over \omega_k'+E_{B''}(k')-W}\,
\nonumber\\ 
    &+& \int\mathrm{d}k'\,
  {{\cal K}_{M'm_\sigma}^{B'\sigma}(k,k')\,
  \widehat{\chi}_{JT}^{\sigma B}(k',k_1,m_\sigma,M) 
         \over \tilde{\omega}_k'+E_{JT}(k')-W}\>,
\label{eq4chiBB}
\end{eqnarray}
with a similar structure of the kernels ${\cal K}_{MM'}^{BB'}(k,k')$
and ${\cal K}_{M'm_\sigma}^{B'\sigma}$.

Equations (\ref{eq4chiN}) and (\ref{eq4chiBB}), as well as 
an analogous consideration for the amplitudes involving the
$\sigma$-meson, imply that the pion amplitudes can be written
in the following form:
\begin{eqnarray}
   \chi_{JT}^{NN}(k,k_0) &=& -\sum_\mathcal{R} c_\mathcal{R}^N(W)
                            \mathcal{V}_{N\mathcal{R}}(k) 
                               +\mathcal{D}^{NN}(k,k_0)\>, \label{cal2chiN}\\
   \widehat{\chi}_{JT}^{B'B}(k,k_1,M',M) &=& 
      -\sum_\mathcal{R}\widehat{c}_\mathcal{R}^B(W,M)
        \mathcal{V}^{M'}_{B' \mathcal{R}}(k) 
               +\mathcal{D}_{M'M}^{B'B}(k,k_1)\>, \label{cal2chiBB}\\
   \widehat{\chi}_{JT}^{B \sigma}(k,k_\mu,M,\mu) &=& 
        -\sum_\mathcal{R}\widehat{c}_\mathcal{R}^\sigma(W,\mu)
         \mathcal{V}^{M}_{B \mathcal{R}}(k) 
               +\mathcal{D}_{M\mu}^{B\sigma}(k,k_\mu)\>, \label{cal2chiBs}\\
   \widehat{\chi}_{JT}^{\sigma B}(k,k_1,\mu,M) &=& 
        -\sum_\mathcal{R} \widehat{c}_\mathcal{R}^B(W,M)
         \mathcal{V}^{\mu}_{\sigma \mathcal{R}}(k) 
               +\mathcal{D}_{\mu M}^{\sigma B}(k,k_1)\>, \label{cal2chisB} \\
   \widehat{\chi}_{JT}^{\sigma\sigma}(k,k_\mu,\mu',\mu) &=& 
        -\sum_\mathcal{R} \widehat{c}_\mathcal{R}^\sigma(W,\mu)
              \mathcal{V}^{\mu'}_{\sigma \mathcal{R}}(k) 
               +\mathcal{D}_{\mu'\mu}^{\sigma\sigma}(k,k_\mu)\>, \label{cal2chiss}
\end{eqnarray}
where $\cal{V}$ are the dressed vertices and $\cal{D}$ are 
the background parts of the amplitudes.

Two simplifications are possible at this point and will 
be only sketched here.  The first one consists in neglecting 
the terms involving the integrals in Eqs.~(\ref{eq4cRN})--(\ref{eq4chiBB}),
resulting in the Born approximation for the $K$ matrix.
In this approximation the $K$ matrix is constructed from 
the meson amplitudes (\ref{cal2chiN})--(\ref{cal2chiss}) 
by using Eqs.~(\ref{chi2KHN}) and (\ref{chi2K}), and replacing
the dressed vertices $\mathcal{V}_{B\mathcal{R}}$ by 
the corresponding bare vertices $V_{B\mathcal{R}}$, 
as well as $\mathcal{D}$ by $\mathcal{K}$.

The second simplification is the averaging over invariant masses,
for instance, the averaging over the $\Delta$ invariant mass $M$
in the $\sigma\Delta$ channel, and invariant mass $\mu$
of the two-pion system occurring in the decay of the $\sigma$ meson.
The averaged invariant masses $\bar{M}$ and $\bar{\mu}$ have been
found by suitable smooth numerical approximations.  In consequence
this type of averaging can be applied to the $K$ matrix and 
the $T$ matrix themselves, and the big advantage of the averaging 
procedure is that the set of integral Heitler equations (\ref{eq4T}) 
becomes a set of algebraic equations. 

In the Figures below the results obtained within the Born
approximation for the $K$ matrix are shown, but we have also
solved the coupled integral equations (\ref{eq4cRN})--(\ref{eq4chiBB})
beyond Born by introducing further approximations, which have allowed
us to write the kernels ${\cal K}^{NN}$, ${\cal K}^{NB}$ and
${\cal K}^{BB'}$ in separable form.  The advantage of using
separable kernels is immense as one is then able to solve
the system exactly.  (Since the quark-$\sigma$ vertex is not 
as well determined as the quark-$\pi$ vertex, the $\sigma$-meson 
vertices have only been treated in the Born approximation.)

Finally, we obtain a set of algebraic equations for the coefficients 
$c^H_\mathcal{R}$, where $H$ stands for $\pi N$, $\pi B$ and $\sigma B$
channels:
$$
\sum_{\mathcal{R}'} A_{\mathcal{R}\mathcal{R}'}(W)c^H_{\mathcal{R}'}(W,m_H) 
   = b_\mathcal{R}^H(m_H)\>,
$$
where
\begin{eqnarray*}
A_{\mathcal{R}\mathcal{R}'} &=&
 (W-M_\mathcal{R}^0)\delta_{\mathcal{R}\mathcal{R}'}
      +   \sum_{B'}
          \int\mathrm{d}k\,{\mathcal{V}^{M_{B'}}_{B'\mathcal{R}}(k)
                              V_{B'\mathcal{R}'}^{M_{B'}}(k)
                        \over \omega_k+E_{B'}(k)-W}\>,
\nonumber \\
b_\mathcal{R}^{B} &=&   V_{B\mathcal{R}}^M(k_1)
       + \sum_{B'} 
           \int\mathrm{d}k\,{\mathcal{D}_{MM_{B'}}^{B'B}(k,k_1)
                       V_{B'\mathcal{R}}^{M_{B'}}(k)
                        \over \omega_k+E_{B'}(k)-W}
         = \mathcal{V}_{B\mathcal{R}}^M(k_1)\>,
\nonumber\\
b_\mathcal{R}^\sigma &=&   V_{N\mathcal{R}}^\mu(k_\mu)\>.
\end{eqnarray*}
Organizing the coefficients and ${\cal V}$ functions into vectors,
${\bi c}^H = (c_\mathcal{R}^H, c_{\mathcal{R}'}^H, \ldots)^T$
and ${\bi {\mathcal{V}}}_H = (\mathcal{V}_{H\mathcal{R}}, 
\mathcal{V}_{H\mathcal{R}'}, \ldots)^T$, 
the solution can be  written in the form
${\bi c}^H = A^{-1}{\bi {\mathcal{V}}}_H$.
The zeros of $A$ occur at the positions of the poles of 
the $K$ matrix related to the resonance $\mathcal{R}$;
the corresponding energies are $M_\mathcal{R}$.
The coefficients $c_\mathcal{R}$ have then been determined
by the following procedure.  First, we have established the zeros 
of the $A$-matrix determinant.  Second, by adjusting the energies 
of the bare states, $M_\mathcal{R}^0$, the poles of the $K$ matrix 
have been forced to acquire some desired values.  Third, 
$A$ has been diagonalized, $UAU^T = D$, such that
$D = \hbox{diag}[\lambda_\mathcal{R}, \lambda_{\mathcal{R}'}, \ldots]
= \hbox{diag}[Z_\mathcal{R}(W)(W-M_\mathcal{R}), 
Z_{\mathcal{R}'}(W)(W-M_{\mathcal{R}'}), \ldots]$,
which defines the wave-function normalization $Z_\mathcal{R}$ 
pertinent to the resonance $\mathcal{R}$.
Finally, this allowed the solution for the $c$ coefficients
to be cast in the form
$$
{\bi c}^H = {U}^T{D}^{-1}{U}{\bi {\mathcal{V}}}_H\>,
$$
while the resonant parts of the $\chi$ amplitudes take the form
\begin{eqnarray*}
   \chi^{H'H} &=&  -{\bi {\mathcal{V}}}_{H'}^T {\bi c}^H      
              =   - {\bi {\mathcal{V}}}_{H'}^T 
                {U}^T{D}^{-1}{U}{\bi {\mathcal{V}}}_H
\nonumber \\
  &=& 
   - \sum_\mathcal{R} \widetilde{\cal V}_{H\mathcal{R}}\,
            {1\over Z_\mathcal{R}(W) (W-M_\mathcal{R})}\,
                    \widetilde{\cal V}_{H'\mathcal{R}}
  =  -\sum_\mathcal{R} \widetilde{c}^{H}_\mathcal{R}
                   \widetilde{\cal V}_{H'\mathcal{R}}\>,
\end{eqnarray*}
where
\begin{equation}
    \widetilde{\cal V}_{H\mathcal{R}} = 
        \sum_{\mathcal{R}'}u_{\mathcal{R}\mathcal{R}'}{\cal V}_{H\mathcal{R}'}\>,
\qquad
    \widetilde{c}_\mathcal{R}^H = {\widetilde{\cal V}_{H\mathcal{R}}
             \over Z_\mathcal{R}(W) (W-M_\mathcal{R})}\>.
\label{calVmix}
\end{equation}
The interpretation of (\ref{calVmix}) is that the resonant states 
$\mathcal{R}$, $\mathcal{R}'$, $\dots$ are not eigenstates of 
the Hamiltonian and therefore mix:
$$
   \widetilde{\Phi}_\mathcal{R} = \sum_{\mathcal{R}'} 
   u_{\mathcal{R}\mathcal{R}'} \Phi_{\mathcal{R}'}\>.
$$

\begin{figure}[!htbp]
\centerline{%
\includegraphics[width=6.3cm]{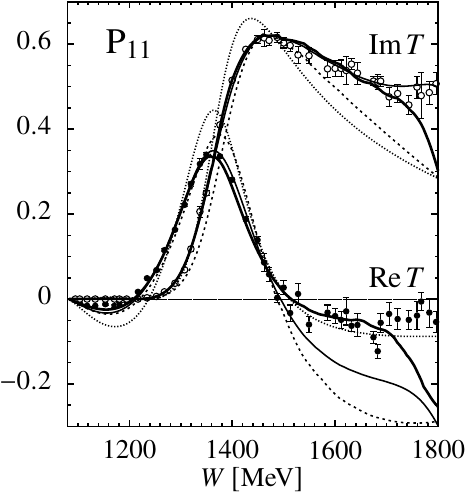}~%
\includegraphics[width=6.3cm]{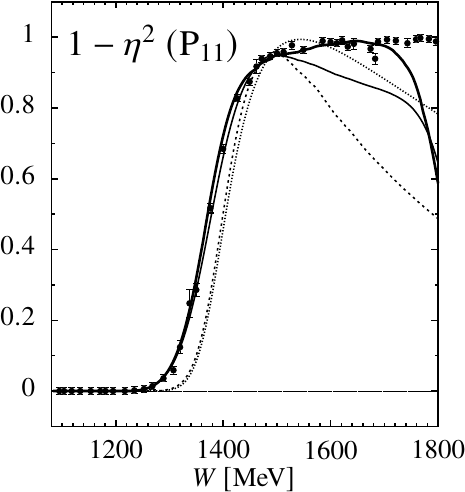}}
\caption{[Left] The real and imaginary parts of the $T$-matrix
for the $P_{11}$ partial amplitude.  Shown are the Born approximation
with resonant terms only (dotted lines), the result including
the background (dashed lines), adding the $\sigma$-channel
(thin solid lines), and the full calculation (thick solid lines).
[Right] The inelasticity in the $P_{11}$ partial wave (same labeling).}
\label{fig:t11}
\end{figure}

In the following the results for the scattering amplitudes in
the $P_{11}$ partial wave are shown.  As mentioned above, we have used
the CBM as the underlying quark model since it is one of the most
popular representations of quark-pion dynamics.  Figure~\ref{fig:t11}
(left) shows the real and imaginary parts of the $T$-matrix,
and Fig.~\ref{fig:t11} (right) shows the inelasticity.

Our study has shown that elastic and inelastic pion-nucleon scattering 
in the energy range from the threshold up to $W\sim 1700\,\mathrm{MeV}$
is governed by an intricate interplay of the $\pi N$, $\pi\Delta$ 
and $\sigma N$ degrees of freedom.  In particular, the correlated 
two-pion decay in the relative $s$-wave by the $\sigma$-meson
turned out to be the crucial ingredient that allowed us to explain
the peculiar feature of the inelasticity in the $P_{11}$ partial wave
just above the two-pion threshold, which quickly increases from zero 
to unity and remains almost constant in a broad energy range.
Our findings also imply that one is able to explain the features 
of the Roper resonance without invoking exotic degrees of freedom, 
and to establish a specific benchmark for an assessment of the 
underlying quark models.  It would be naive to expect, however, 
that the scattering analysis alone would be able to provide 
a definitive criterion for feasible models, so we next turned 
to the application of the coupled-channel method to full 
electro-production amplitudes.

\subsection{Calculation of photo- and electro-production amplitudes}

Photo- and electro-production have been incorporated into 
the coupled-channels $K$-matrix formalism by including
a new channel, $\gamma N$.  Because the electro-magnetic 
interaction is considerably weaker than the strong interaction, 
we have assumed
$K_{\gamma N\,\gamma N}\ll K_{\gamma N\,\pi N}\ll K_{\pi N\,\pi N}$,
and similarly for other channels.  The electro-magnetic interaction
Hamiltonian has been taken in the form
$$
  H_\gamma ={1\over\sqrt{2\pi}^3}\int\mathrm{d}\vec{k}_\gamma  \sum_\mu\,
   \left[\tilde{V}^\gamma_\mu(\vec{k}_\gamma)a_\mu(\vec{k}_\gamma) 
                  + \hbox{h.c.}\right]\,,
$$
where  $\vec{k}_\gamma$ and $\mu$ are the momentum 
and the polarization of the incident photon, and
\begin{equation}
\tilde{V}^\gamma_\mu(\vec{k}_\gamma)
  = {\mathrm{e}_0\over\sqrt{2\omega_\gamma}}
    \int\mathrm{d}\vec{r}\,\vec{\varepsilon}_\mu\cdot\vec{j}(\vec{r})
    \,\e^{\mathrm{i}\vec{k}_\gamma\cdot\vec{r}} \>.
\label{Vgamma}
\end{equation}
The $K$-matrix elements for electro-production corresponding to 
different channels $MB$ ($\pi N, \pi\Delta$, $\sigma N, \ldots$) 
are introduced as expectation values of (\ref{Vgamma}) between
the state representing the photon-nucleon system, $\Psi_N$,
and the principal-value states,
$$
\mathcal{M}^{K\,{JI}}_{MB}
  = -{\mathcal{N}_\gamma\over \sqrt{k_0 k_\gamma}}\,
     \langle{\Psi}_{JI}^{MB}(m_Jm_I;k_0,l)|
      \tilde{V}^\gamma_\mu(\vec{k}_\gamma)|\Psi_N(m_s m_t)
  \rangle \>,
$$
where $J$ and $I$ denote spin and isospin, respectively.
They are related to the electro-production amplitudes through 
$\mathcal{M}=\mathcal{M}^K+\mathrm{i}\, T\mathcal{M}^K$, a relation 
that follows from the Heitler equation, Eq.~(\ref{TSmatrix}).
For the $P_{11}$ partial wave in the region of the Roper resonance 
it turned out that only the $\pi\Delta$ and the $\sigma N$ 
inelastic channels are needed, and we end up with \cite{Golli:2009uk}
\begin{eqnarray}
\mathcal{M}_{\pi N}(W) &=& \mathcal{M}_{\pi N}^K(W)
      +\mathrm{i}\biggl[T_{\pi N\pi N}(W)\mathcal{M}_{\pi N}^K(W)
\biggr. \nonumber \\ &+& \biggl.
       \int_{M_N+m_\pi}^{W-m_\pi} \kern-3pt\mathrm{d} M_\Delta \,
           T_{\pi N\pi\Delta}(W,M_\Delta)
        \mathcal{M}_{\pi\Delta}^K(W,M_\Delta)
\biggr. \nonumber \\ &+& \biggl.
       \int_{2m_\pi}^{W-M_N} \kern-3pt\mathrm{d} \mu 
          \,T_{\pi N\sigma N}(W,\mu)
            \mathcal{M}_{\sigma N}^K(W,\mu)
\biggr]\>.
\label{MgamNpiNa}
\end{eqnarray}
Close to a resonance $\mathcal{R}$ the $K$-matrix element between 
the elastic channel and an arbitrary channel $MB$ can be split 
in the resonant and the background parts,
$$
    K_{\pi N\, MB} = 
  -\pi\sqrt{\omega_0\omega_ME_NE_B\over k_0k_MW^2}
   \, \widetilde{c}_{\mathcal{R}}^{MB}
    \widetilde{\mathcal{V}}^\pi_{N{\mathcal{R}}}(k_0) 
     + K_{\pi N\,MB}^{\mathrm{bg}} \>,
$$
and this allows us to split the electro-production amplitudes
in the same fashion: the resonant part takes the form
$$
\mathcal{M}_{\pi N}^\mathrm{(res)}  = 
    -\sqrt{\omega_\gamma E_N^\gamma\over\pi^2\omega_0E_N}
      {\sqrt{Z_{\mathcal{R}}}\over \mathcal{V}_{N{\mathcal{R}}}}
\,\langle\widehat{\Psi}_{\mathcal{R}}^\mathrm{(res)}(W)|\tilde{V}^\gamma
                 |\Psi_N\rangle\, T_{\pi N\pi N}\,,
$$
while the background part satisfies
$$
\mathcal{M}_{\pi N}^\mathrm{(bg)} = 
       \mathcal{M}_{\pi N}^{K\,\mathrm{(bg)}}
      +\mathrm{i}\biggl[T_{\pi N\pi N}\mathcal{M}_{\pi N}^{K\,\mathrm{(bg)}} 
+ \overline{T}_{\pi N\pi\Delta}
  \overline{\mathcal{M}}_{\pi\Delta}^{K\,\mathrm{(bg)}} 
+ \overline{T}_{\pi N\sigma N}
  \overline{\mathcal{M}}_{\sigma N}^{K\,\mathrm{(bg)}}
\biggr] \>.
$$
The remaining steps are the multipole decomposition and the
calculation of helicity amplitudes.  Omitting all details,
let us just write down as an illustration a sample relation 
between a specific electro-production and the helicity amplitude.
We obtain
\begin{eqnarray}
{\rm Im}\,{}_pM_{1-}^{(1/2)} &=& 
     -{1\over3}\sqrt{k_W E_N^\gamma\over6\pi^2\omega_0E_N}
      {\sqrt{Z_{\mathcal{R}}}\over \mathcal{V}^\pi_{N\mathcal{R}}}
\;{\rm Im} T_{\pi N\pi N}
\left(-{3\over\sqrt2}\right) \nonumber \\
&&\times
\left\langle \widehat{\Psi}_{\mathcal{R}}^\mathrm{(res)}\,
\left(m_s=\half\right)\left|\tilde{V}^{M1}\right|\Psi_N\, 
\left(m_s=-\half\right) \right\rangle \>,
\label{ImpMm12}
\end{eqnarray}
where
$$
   \Gamma_{\pi N} = 
          2\pi {\omega_0 E_N\mathcal{V}^\pi_{N\mathcal{R}}(k_0)^2
                  \over Z_{\mathcal{R}}k_0W}
$$
is the elastic width of the resonance.  We thus obtain
$$
 {\rm Im}\,{}_pM_{1-}^{(1/2)} = -\xi_{\mathcal{R}} 
     \sqrt{k_W E^\gamma_N\Gamma_{\pi N} \over
      6\pi k_0 M_{\mathcal{R}} \Gamma^2}\; A^p_{1/2} \>.
$$
Phenomenological studies of electro-production reveal a relatively
strong contribution of the $\omega$-meson already at low energies,
whereas in the calculation of the (strong) scattering amplitudes
the contributions of vector mesons are negligible.  The contribution
of the $\omega$-meson to the $K$-matrix in the elastic channel 
has therefore been modeled by the expression
$$
{}_pM^{(1/2)}_{1-}(\omega\hbox{-meson})
= {1\over3}\; {M_N\over 4\pi Wm_\pi}\,
{ g_{\gamma\pi\omega}g_{\omega 1}\,k_\gamma k_\pi\,
      \rho_\omega(k_\omega)     \over 
  m_\omega^2 -m_\pi^2 + 2k_\gamma\omega_\pi} \>,
$$
where $\rho_\omega({k}_\omega)$ is the appropriate form-factor;
see \cite{Golli:2009uk} for details.

\goodbreak

\begin{figure}[!htbp]
\centerline{%
\includegraphics[width=6.3cm]{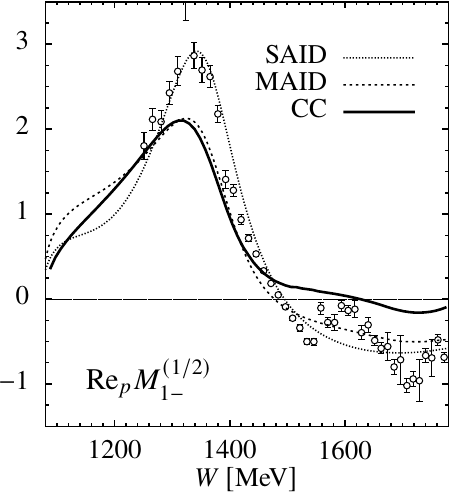}~%
\includegraphics[width=6.3cm]{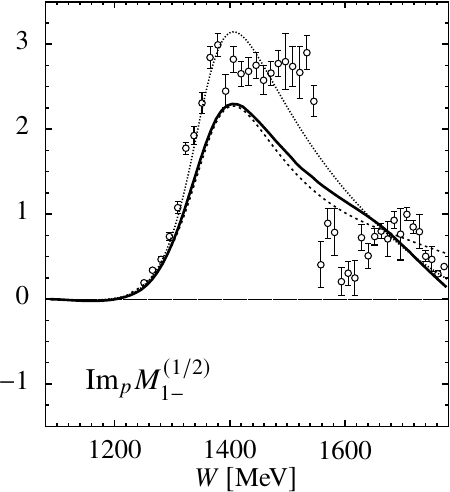}}
\caption{[Left] Real and [Right] imaginary part (see Eq.~(\ref{ImpMm12}))
of the ${}_pM_{1-}^{(1/2)}$ photo-production amplitude.  The experimental 
points are the single-energy solutions of the SAID partial-wave analysis 
\cite{PhysRevC.52.2120,PhysRevC.69.035213}; the ``SAID'' curve shows 
the corresponding fit; the MAID result is based on the parameterization
provided by \cite{MAID}.}
\label{fig:pM1m}
\end{figure}

Figure~\ref{fig:pM1m} shows a sample result for the ${}_pM_{1-}^{(1/2)}$
amplitude by using the same parameters for the CBM model as in the
calculation of the scattering amplitudes.  One of the more interesting
findings in the $P_{11}$ case was that at energies below the resonance 
the amplitudes are dominated by the background, which is in marked 
contrast to the $P_{33}$ partial wave in the region of the $\Delta(1232)$,
where the photo-production amplitude is clearly dominated by 
the resonant contribution and follows the shape of the elastic $T$ matrix.

\begin{figure}[!htbp]
\centerline{%
\includegraphics[width=7cm]{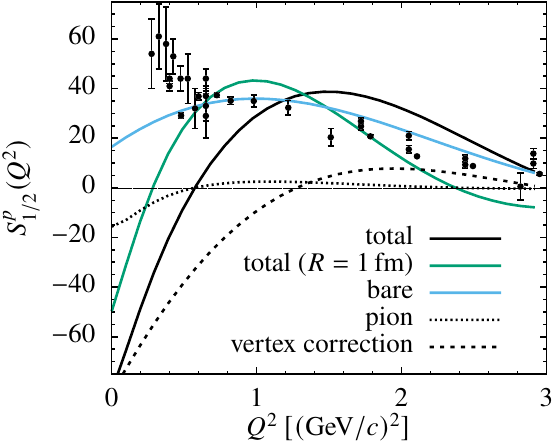}}
\caption{Scalar helicity amplitude $S_{1/2}^p(Q^2)$ evaluated at 
the pole of the $K$-matrix ($W=1520\,\mathrm{MeV}$).  Separately shown 
are the contributions of the $3q$ core (``bare''), the $\gamma\pi\pi'$
interaction (``pion''), and the pion-cloud corrections to the
$\gamma BB'$ vertex (``vertex correction'').  An additional ``total'' 
curve is plotted for a non-standard bag radius of $1\,\mathrm{fm}$
(otherwise $0.83\,\mathrm{fm}$).}
\label{fig:Sh1520}
\end{figure}

As a further example, Fig.~\ref{fig:Sh1520} shows the scalar
helicity amplitude, evaluated at $W=1520\,\mathrm{MeV}$,
the pole of the $K$-matrix.  We observed an interesting 
feature that the effects of the pion cloud which,
understandably, are most pronounced at small $Q^2$, 
have the opposite sign with respect to the contribution 
from the quark core, which is small at the origin---with 
the same mechanism also governing the behavior 
of the magnetic helicity amplitude (not shown).
The large spread of various quark-model predictions
for the scalar helicity amplitude near the real-photon point 
and the uncertainties of the helicity couplings $S_{1/2}^p(0)$
found by SAID and MAID motivated us for an experimental
exploration mentioned in Section~\ref{sec:exp}.

\subsection{Including the ``second Roper''}

An additional impetus to investigate the structure of the Roper 
resonance came as part of a larger enterprise to study eta and kaon 
photo-production on nucleons within the same coupled-channel approach 
as described above.  The required energy range necessarily extended
up to $\approx 1800$~MeV, and this naturally led to the inclusion
of the ``second Roper'', the $N^\ast(1710)$, for which we assumed 
that it decays only into the $\sigma N$ channel.  Furthermore, production
of strange mesons in the $P_{11}$ partial wave required us to specify 
its quark configuration, which is not obvious;
it may consist of the $(1s)^3$, $(1s)^2(2s)^1$, $(1s)^1(2s)^2$,
$(1s)^1(1p)^2$, $(1s)^1(1d)^2\ldots$ configurations or even a 
component with the excitation of the $\sigma$ cloud \cite{Alberto:2001fy}.
We have restricted our investigation to the type of excitation in which
only a single $1s$ quark is promoted to a higher orbit, which resulted
in the following structure of the two Roper resonances \cite{Golli:2016dlj}:
\begin{eqnarray*}
N^\ast(1440) &=& \cos\vartheta_R(1s)^2(2s)^1 - \sin\vartheta_R(1s)^1(1x)^2 \>, \\
N^\ast(1710) &=& \sin\vartheta_R(1s)^2(2s)^1 + \cos\vartheta_R(1s)^1(1x)^2 \>.
\end{eqnarray*}
Here $x$ stands for $l>0$ quark orbits not involved in the
transition matrix elements.  Note that the $N^\ast(1440)$ 
and the ground state also mix, which leads to an important 
nucleon-pole contribution to the amplitudes.

\subsection{Genuine quark state versus dynamically generated structure}

Our last theoretical effort was motivated by recent lattice QCD 
studies (Graz, Adelaide) that have found no clear signal for 
a dominant three-quark configuration below $1.65\,\mathrm{GeV}$ 
and $2.0\,\mathrm{GeV}$,
respectively, that could be interpreted as a Roper state.
It seemed that the $\pi N$ channel alone does not render 
a low-lying resonance and that coupling with $\pi\pi N$ channels 
seems to be important, supporting the dynamical origin of the Roper.
With our coupled-channels machinery in place, we decided to investigate
whether such a dynamical mechanism for the formation of the Roper 
would be possible.  This work has been performed in collaboration 
with Tuzla (H.~Osmanovi\'c) and Zagreb (A.~\v{S}varc) groups
\cite{Golli:2017nid}.

The pion-baryon vertices in the underlying quark model (still CBM)
have been fixed, while the $s$-wave sigma-baryon interaction 
has been incorporated phenomenologically with the coupling strength, 
the mass and the width of the $\sigma$ meson introduced as free parameters.
The Laurent-Pietarinen expansion has been used to extract 
the parameters of the $S$-matrix pole.  The Lippmann-Schwinger 
equation for the $K$ matrix with a separable kernel has been solved 
to all orders.  The main finding was that for sufficiently strong 
$\sigma NN$ coupling the kernel becomes singular and a quasi-bound 
state emerges at around $1.4\,\mathrm{GeV}$, dominated by the $\sigma N$ 
component and manifesting itself as a pole of the $S$-matrix.
As an alternative we have added a $(1s)^2 2s$ quark resonant state, 
and studied the interplay of the dynamically generated state and 
the three-quark resonant state.  It turned out that for the mass 
of the three-quark resonant state above $1.6\,\mathrm{GeV}$ 
the mass of the resonance is indeed determined solely by 
the dynamically generated state, yet with the caveat that 
the three-quark resonant state remains imperative to reproduce 
the experimental width and the modulus of the resonance pole.

\section{Experimental studies of the Roper resonance}
\label{sec:exp}

\begin{figure}[!htbp]
\centerline{%
\includegraphics[width=8cm]{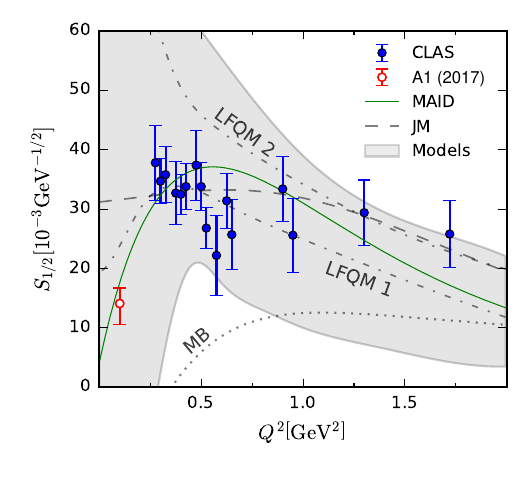}}
\caption{The scalar helicity amplitude for Roper electro-excitation 
at $Q^2 = 0.1 \, (\mathrm{GeV}/c)^2$, denoted by ``A1 (2017)'',
compared to CLAS data, MAID, the JLab-MSU parameterization,
and two light-front quark model results.  The ``MB'' curve shows
the meson-baryon dressing contribution.  The immense range of 
various model predictions is indicated by shading; see \cite{Stajner:2017fmh}
for detailed references.}
\label{fig:PRLs12}
\end{figure}

The Ljubljana group also led an experimental effort within
the A1 Collaboration at the MAMI facility.  We have precisely
measured the proton recoil polarization components in the
$p(\vec{e},e' \vec{p})\pi^0$ process in the energy range
of the Roper resonance and, by using the MAID unitary isobar 
model, have been able to determine (in a model-dependent
manner) the scalar helicity  amplitude $S_{1/2}$ at a $Q^2$ 
very close to the real-photon point \cite{Stajner:2017fmh}.  
In view of the cancellations of bare-core and pion-cloud 
contributions seen by some models at low $Q^2$, this region 
is relevant as a kind of ``elimination ground'' for quark models.  
Our extracted value (see Fig.~\ref{fig:PRLs12}) is consistent 
with the non-hybrid nature of the resonance, which implies that
the interpretations of the Roper as an entity characterized 
by strong meson-baryon dressing may be favored, in other words,
there is no need for non-quark degrees of freedom.  Our result
therefore supports those models of the Roper in which the interplay 
of quark and meson contributions results in a small value of $S_{1/2}$
near the real-photon point.

\section*{Acknowledgment}

With this paper the Author would like to pay tribute to late 
Professor Bojan Golli who for many years was the driving force 
in nucleon resonance-related theoretical efforts of the Ljubljana group.

\bibliographystyle{unsrt}
\bibliography{main}

\end{document}